# Potentiometric detection of spin polarization expected at the surface of FeTe$_{0.6}$Se$_{0.4}$ in the effective *p*-wave superconducting state

Kosuke Ohnishi [1,$], Ryo Ohshima [1,#,$], Taiki Nishijima [1], Shinya Kawabata [1],

Shigeru Kasahara [2,3], Yuichi Kasahara [3], Yuichiro Ando [1,4],

Yoichi Yanase [3], Yuji Matsuda [3], and Masashi Shiraishi [1,#,$]

1. Department of Electronic Science and Engineering, Kyoto University, Kyoto 615-8510, Japan.

2. Research Institute for Interdisciplinary Science, Okayama University, Okayama, 700-8530, Japan.

3. Department of Physics, Kyoto University, Kyoto 606-8502, Japan

4. PRESTO-JST, Honcho, Kawaguchi, Saitama 332-0012, Japan.

# Corresponding authors: Masashi Shiraishi (shiraishi.masashi.4w@kyoto-u.ac.jp) and Ryo Ohshima (ohshima.ryo.2x@kyoto-u.ac.jp)

$ These authors equally contributed to the work.

**Nowadays, the quest for non-Abelian anyons is attracting tremendous attention. In particular, a Majorana quasiparticle has attracted great interest since the non-Abelian anyon is a key particle for topological quantum computation. Much effort has been paid for the quest of the Majorana state in solids, and some candidate material platforms are reported. Among various materials that can host the Majorana state, chiral *p*-wave superconductor is one of the suitable materials and the iron-based layered superconductor FeTeSe is one of the promising material platforms because its surface can host effective *p*-wave superconducting state that is analogous to chiral *p*-wave superconducting state thanks to its topological surface state. Given that a chiral**

**$p$-wave superconductor possesses spin polarization, detecting the spin polarization can be evidence for the chiral $p$-wave trait, which results in the existence of Majorana excitation. Here, we show successful detection of the spin polarization at the surface of $FeTe_{0.6}Se_{0.4}$ in its superconducting state, where the spin polarization is detected via a potentiometric method. Amplitudes of the spin signal exhibit characteristic dependence for temperature and bias current, suggesting detection of spin polarization of the Bogoliubov quasiparticles. Our achievement opens a new avenue to explore topological superconductivity for fault-tolerant quantum computation.**

A quest of chiral $p$-wave superconductors [1], a form of a topological superconducting state, has been collecting great attention in condensed matter physics because of significance to find a new family of unconventional superconductors and to explore a new material platform hosting the Majorana quasiparticles enabling the fault-tolerant quantum computing [2]. In addition to a chiral $p$-wave superconductor, the theoretical investigation by Fu and Kane [3] allows prediction of the existence of an effective $p$-wave superconducting state in an iron-chalcogenide high-$T_c$ superconductor, e.g., FeTeSe. The superconducting state of FeTeSe can be regarded as the $p$-wave superconducting state under basis transformation, resulting in possible hosting of the Majorana state, although its bulk state is an $s$-wave superconductor in the conventional basis. Indeed, recent studies using scanning tunnelling microscopy studies revealed that the Majorana bound state can appear in vortex cores in FeTeSe [4,5] with the assistance of a perpendicular external magnetic field, where the zero-energy state in the vortex core that is attributable to the Majorana bound state was observed. The observation of the zero-energy state can be evidence that the surface of FeTeSe possesses the effective topological *p*-wave superconducting nature. The other significant studies are the detection of spontaneous time-reversal symmetry breaking in FeTeSe by means of angle-resolved photoemission spectroscopy (ARPES) and magnetic flux measurements using nitrogen-vacancy (NV) centers in diamond and the Kerr effect [6-8], where the energy-gap-opening at the Dirac point due to the symmetry breaking, detection of an effective magnetic field and discernible Kerr rotation at the surface of FeTeSe were observed, respectively. Albeit some experimental challenge to pursue and claim the possible *p*-wave nature have been so far reported, the experimental procedures to enable accessing the effective *p*-wave nature are still limited, which hampers creation of electrically driven devices using FeTeSe that can allow Majorana quasiparticle detection and utilization. As clarified in theory [9], chiral *p*-wave superconductors can host spin polarization even in excited quasiparticles, which can take place in FeTeSe. Meanwhile, detection of the spin polarization in the effective *p*-wave states is as yet unsuccessful, whereas the *p*-wave superconducting state hosting Majorana quasiparticles can open a novel and necessary pathway for future fabrication of topological computing devices. Given that evidence of Majorana quasiparticles due to chiral *p*-wave superconductivity in solid devices is still under debate, fabricating nano-sized electronic/spintronic devices using a chiral *p*-wave superconductor is a major and invaluable challenge in both fundamental topological physics and device physics for nanoelectronics. In this paper, we demonstrate direct potentiometric (electric) detection of spin polarization in the effectively chiral *p*-wave superconducting state at the surface of $FeTe_{0.6}Se_{0.4}$.

An $FeTe_{0.6}Se_{0.4}$ bulk single crystal was synthesized using the chemical vapor transport method, and an $FeTe_{0.6}Se_{0.4}$ thin film was mechanically exfoliated and transferred onto a $SiO_2$(100 nm)/Si substrate using the scotch tape method. Ar+ milling was implemented to remove a possible oxidized layer on the FeTe$_{0.6}$Se$_{0.4}$ surface. The thickness of the FeTe$_{0.6}$Se$_{0.4}$ was about 70 nm, and Au(70 nm)/Co(30 nm) electrodes are equipped onto the FeTe$_{0.6}$Se$_{0.4}$ as a spin detector, where the Co was deposited by using electron beam deposition. Device fabrication was carried out by using electron beam lithography. The resistance measurements to check the superconductivity of FeTe$_{0.6}$Se$_{0.4}$ and NbN were carried out by using the conventional four-probe method. For the spin voltage measurements, in addition to resistance measurements, a closed-cycle cryostat system (Niki Glass Co., LTD) and a physical property measurement system (Quantum Design) were used.

Figures 1(a) and 1(b) show a schematic of a spin device using FeTe$_{0.6}$Se$_{0.4}$ and its optical microscopy image, respectively. The thickness of the FeTe$_{0.6}$Se$_{0.4}$ was several tens of nanometers (see Methods for the retail). The accomplished potentiometric method is applied, in which the middle Co electrode plays the role of a spin detector, i.e., the local electrical three-terminal method [10-14]. The method allows efficient detection of spin voltages as the high- to the low-states by changing spin alignments (parallel and anti-parallel) between spins in the material and the ferromagnetic electrode. Although spin polarization of Cooper pairs cannot be detected in a potentiometric method because the electrochemical potentials of Cooper pairs cannot be defined due to their nondissipative trait, the method is applicable to the detection of the electrochemical potential of Bogoliubov quasiparticles, the electrochemical potential of which can be well defined due to its dissipative nature. More importantly, the spin polarization attributed to the effectively chiral *p*-wave nature of Cooper pairs can be transformed to that of Bogoliubov quasiparticles [9], and thus, the electrochemical potential of the spin polarization of Bogoliubov quasiparticles is detected by controlling the spin direction of the Co electrode by sweeping an in-plane external magnetic field.

The superconducting transition of FeTe$_{0.6}$Se$_{0.4}$ was confirmed by measuring a steep decrease in the resistance (see Fig. 1(c)), and the superconducting transition was observed at approximately 11 K using the midpoint of the normal-state resistance (defined as $T_c$), which is almost comparable to that in previous studies (approximately 14 K).

Figures 1(d) and 1(e) are the central results of this work, where salient spin voltage hysteresis can be seen at 7 K (below $T_c$; see Fig. 1(d)), whereas the hysteresis almost disappears at 16 K (above $T_c$; see Fig. 1e). The small decrease in the spin signals at $\pm 25$ mT observed at 16 K is attributed to the anisotropic magnetoresistance of the single Co electrode, which is usually observed in such a local electrical spin detection method and is evidence for the magnetization reversal, i.e., the coercive force of the Co electrode [12,14]. The injection electric current was set to 10 µA, which is below the critical current. The observation of the noticeable voltage hysteresis at 7 K is reminiscent of successful detection of spin polarization due to the spin polarization attributed to the effectively chiral $p$-wave trait at the surface of FeTe$_{0.6}$Se$_{0.4}$ in a superconducting state, since the hysteresis loop width is consistent with the coercive force of the Co electrode. To corroborate the electrical detection of the spin polarization of the Bogoliubov quasiparticles, the temperature dependence of the spin voltages $\Delta V$ was measured, where $\Delta V$ is defined as the difference in the spin voltages at 0 mT under upward and downward sweeping of the magnetic field considering possible thermal drift in the output voltages (see also Supplemental Material [15]). Figure 1(f) shows the whole dataset of the measured $\Delta V$ as a function of temperature. When the resistance of FeTe$_{0.6}$Se$_{0.4}$ starts to decrease from 16 K, the superconducting gap partly appear in the FeTe$_{0.6}$Se$_{0.4}$, where the intermediate state hosting coexisting superconducting and normal state regions in the FeTe$_{0.6}$Se$_{0.4}$ appears, and spin voltages can be detected due to the local superconducting region. As the resistance of FeTe$_{0.6}$Se$_{0.4}$ decreases, $\Delta V$ almost monotonically increases and reaches the maximum at 6-8 K, where the region with the effectively chiral *p*-wave superconductivity nature expands and the superconducting state of FeTe$_{0.6}$Se$_{0.4}$ is dominant (see also Fig. 1(c)), where the superconducting gap fully open in the FeTe$_{0.6}$Se$_{0.4}$. The amplitude of $\Delta V$ then decreases until 4 K since the expansion of the superconducting gap suppresses excitation of the Bogoliubov quasiparticles, which is in principle consistent with the gap opening at the surface of superconducting FeTe$_{0.6}$Se$_{0.4}$ confirmed by scanning tunnelling microscopy [5]. We also note that a previous study corroborated that spin polarization in a spin current can be transferred to Bogoliubov quasiparticles and the quasiparticles can carry the spin polarization [16]. Thus, all results can be comprehensively understood under the prerequisite that the effectively chiral *p*-wave superconducting state possessing spin polarization of Cooper pairs appears at the surface of FeTe$_{0.6}$Se$_{0.4}$ and the spin polarization is transformed to Bogoliubov quasiparticles.

To obtain further compelling evidence, the following control experiments were implemented: Figures 2(a) and 2(b) show the measurement setups and observed spin voltages. The injection electric current was set to 10 μA, the same as in the measurements shown in Fig. 1. In addition to the electrical three-terminal measurement using a ferromagnetic electrode, the three-terminal measurement using a nonmagnetic electrode and the conventional four-terminal measurement were implemented to negate possible detection of artefacts in spin voltages that could occur in the three-terminal measurement [17-19]. Whereas no voltage hysteresis was observed when a nonmagnetic electrode was used as a detector (Fig. 2(a)), noticeable hysteresis in the spin voltages was observed in the four-terminal setup when using a Co electrode (Fig. 2(b)). The findings rationalize successful detection of the spin polarization due to the effectively chiral *p*-wave state in FeTe$_{0.6}$Se$_{0.4}$, and importantly, the spin voltage hysteresis is not attributed to possible artefacts reported in the previous works [17-19] because the spin voltage hysteresis is always detected in all measurements when the Co electrode is used as a spin detector. We also implemented the other experiment, where the sample was cooled down with applying an external magnetic field (i.e., the field cooling experiment) to negate possible contribution of emergent magnetic moments in the FeTe$_{0.6}$Se$_{0.4}$ channel [7,8,20] (see Supplemental Material [15]).

To negate unwanted formation of the secondary phases between the Co and the FeTe$_{0.6}$Se$_{0.4}$, which may give rise to spin signals in the measurements, we prepared a similar spin device made of FeSe. FeSe is the mother superconducting material of FeTe$_{0.6}$Se$_{0.4}$, possessing a very similar crystal structure and layered structure, and is a conventional $s$-wave superconductor unlike FeTe$_{0.6}$Se$_{0.4}$. If the secondary phase layer allows creation of spin polarization, the same spin voltage hysteresis could be observed. More importantly, this control experiment has significance to substantiate that the effective $p$-wave trait enables spin polarization given that FeSe is a conventional $s$-wave superconductor. The schematic image of the device and the results are shown in Figs. 3(a) and (b), where the spin voltage from the FeSe is indiscernible above and even below its superconducting transition temperature (see also Supplemental Material [15] for the detail). Given that the device fabrication process of the FeSe and FeTe$_{0.6}$Se$_{0.4}$ devices are completely the same, this result unequivocally negates that the secondary phase formation might induce unwanted spin polarization in the measurement using the FeTe$_{0.6}$Se$_{0.4}$ and signifies as well that the effective $p$-wave trait of FeTe$_{0.6}$Se$_{0.4}$ allows spin signal creation. Regarding the missing spin polarization in $s$-wave superconductors, a spin voltage measurement using the other conventional $s$-wave superconductor, NbN, provides supporting evidence. Figure 3(c) shows a schematic image of the NbN device, where the thicknesses of the NbN and the Au/Co spin detector were set to be 30 nm and 80/20 nm, respectively. Figures 3(d)-3(f) show the resistance of NbN as a function of temperature and the spin voltages as a function of the external magnetic field at 4 K and 6 K. Although a superconducting transition is observed in NbN below 5 K, no hysteresis in the spin voltages can be seen below and above 5 K (see Figs. 3(e) and 3(f)). The absence of the spin voltage hysteresis shown in Fig. 3(e) can be rationalized by the lack of spin polarization in the *s*-wave superconducting state, and the absence also indicates that a possible spin-orbit coupling induced by the FM does not allow spin triplet formation. Simultaneously, the results obtained using NbN underscore the validity of potentiometric detection of the effectively chiral *p*-wave superconducting state at the surface of FeTe$_{0.6}$Se$_{0.4}$, although an origin the voltage difference in the positive and the negative external magnetic fields is unclear. Here, we note once again that the lack of the spin voltage hysteresis is the key to negate the spin triplet state in NbN as discussed in the literature [21] (see also Supplemental Material [15] for dependence of the three-terminal spin voltages on the angle of the external magnetic field in the FeTe$_{0.6}$Se$_{0.4}$ and NbN devices).

The spin voltage $\Delta V$ is corroborated to be observable below the transition temperature in the previous paragraphs. To demarcate the observable domain in terms of the amplitude of the injected electric current, the current amplitude dependence of the spin signals was measured by preparing another FeTe$_{0.6}$Se$_{0.4}$ device. Figures 4(a) and 4(b) show the spin voltages in a difference device from that exhibiting the results shown in Fig. 1 measured with currents $I = +50$ μA and $-50$ μA, respectively. Spin voltage hysteresis due to the magnetization reversal of the Co electrode as aforementioned is observed in each measurement condition, where the polarity of the hysteresis is unchanged, which is reproducible in the other FeTe$_{0.6}$Se$_{0.4}$ device (see Supplemental Material [15]). Albeit its underlying physics is somewhat elusive, we may resort to the results on time-reversal symmetry breaking in Fe-based superconductor that allows preserving spin polarization to interpret our result [6-9].

Furthermore, it is also noteworthy that a conventional *s*-wave superconductor enables the sign reversal of the spin voltage hysteresis [22], which is not consistent with our result. Hence, the unchanged polarity of the hysteresis is ascribable to an unconventional superconducting nature of $FeTe_{0.6}Se_{0.4}$. Spintronic heating effects, such as spin-dependent Seebeck or anomalous Nernst effects, are eliminated as an origin of the spin signals as well (although they may allow unchanged polarity of the sign), since the spin voltages are observable only from $FeTe_{0.6}Se_{0.4}$ whereas the ferromagnetic electrodes in the $FeTe_{0.6}Se_{0.4}$, FeSe and NbN devices are identical. In addition, the bias current dependence of the spin signals (see Fig. 4(c) shown in the next paragraph) is not accountable by the heating effects because the amplitude of the spin signals does not monotonically increase as a function of the current amplitude. Meanwhile, it is noteworthy that further detailed theoretical calculation is necessary to precisely clarify how the chiral spin texture appears in the Bogoliubov quasiparticles because the order parameter itself is not the chiral *p*-wave but it behaves as effectively chiral *p*-wave one after the basis transformation in the case of $FeTe_{0.6}Se_{0.4}$. Thus, further investigation by a combination of theory and experiments is requested.

Figure 4(c) shows the evolution of the spin voltage hysteresis as a function of the amplitude of the injected electric current. Whilst the amplitude of the spin voltage gradually increases up to 50 μA, it monotonically decreases under larger current amplitudes (Fig. 4(c)). The decrease of the spin voltage above 50 μA is ascribable to instability of superconductivity because the current amplitude approaches to the critical current. In fact, the spin voltage hysteresis disappears and only anisotropic magnetoresistance (AMR) signals due to the magnetization reversal of the Co electrode of 30 nm in thick are observed when $I$ = 200 μA (Fig. 4(d)). Since the critical current of the $FeTe_{0.6}Se_{0.4}$ was measured to be ca. 200 μA (see Supplemental Material [15]), the absence of the spin voltage hysteresis at $I \geq 200$ µA underpins our assertion that the spin voltage hysteresis observed from the FeTe$_{0.6}$Se$_{0.4}$ spin devices is ascribed to the spin polarization in the FeTe$_{0.6}$Se$_{0.4}$ (i.e., the breaking of the superconductivity of the FeTe$_{0.6}$Se$_{0.4}$ gives rise to the disappearance of the spin voltage hysteresis). Figure 4(e) shows the whole trend of the amplitudes of the spin signals, and notably, a similar trend is observed under a negative current (see Supplemental Material [15]). The comprehensive understanding of the whole results is as follows: Whereas the total number of excited Bogoliubov quasiparticles increases as the current amplitude increases, the superconducting state of FeTe$_{0.6}$Se$_{0.4}$ becomes unstable when the current amplitude approaches the critical current, and finally, the superconductivity diminishes, resulting in the disappearance of spin-polarized quasiparticles.

Spin polarization that appears in the effectively chiral *p*-wave superconducting state hosting Majorana quasiparticles was electrically detected through a combination of a layered superconductor, FeTe$_{0.6}$Se$_{0.4}$, and the established potentiometric method. The spin voltage ascribed to successful detection of spin polarization of Bogoliubov quasiparticles was the maximum when the superconducting gap started to open and monotonically decreased as the gap expanded, whereas the voltage disappeared above the superconducting transition temperature and above the critical electric current. The absence of spin voltages in a conventional *s*-wave superconductor, NbN, strongly underscores the validity of our assertion. This achievement in this work can pave a novel way not only to search for topological superconductors available for fault-tolerant quantum computing using Majorana quasiparticles but also to investigate spin-polarized states appearing in a wide variety of superconductors and possible spin-triplet states appearing in heterostructure systems possessing substantial SOI.

**Acknowledgements**

This research is supported in part by the Japan Society for the Promotion of Science (JSPS) Grant-in-Aid for Challenging Research (Pioneering) (No. 20K20443 and No. 21K18145) and JSPS Grant-in-Aid for Scientific Research (No. 22H01181 and No. 22H04933). The authors (K.O., R.O., Y.A. and M.S.) are grateful to Dr. E. Tamura and Prof. Y. Suzuki of Osaka Univ. for fruitful discussion.

**References**

(1) M. Sato and Y. Ando, Topological superconductors: A review, Rep. Prog. Phys. **80**, 076501 (2017).

(2) A. Kitaev, Anyons in an exactly solved model and beyond, Ann. Phys. **321**, 2-111 (2006).

(3) L. Fu and C. L. Kane, Superconducting Proximity Effect and Majorana Fermions at the Surface of a Topological Insulator, Phys. Rev. Lett. **100**, 096407 (2008).

(4) D. Wang, L. Kong, P. Fan, F. Chen, S. Zhu, W. Liu, L. Cao, Y. Sun, S. Du, J. Schneeloch, R. Zhong, G. Gu, L. Fu, H. Ding, and H.-J. Gao, Evidence for Majorana bound states in an iron-based superconductor, Science **362**, 333 (2018).

(5) T. Machida, Y. Sun, S. Pyon, S. Takeda, Y. Kohsaka, T. Hanaguri, T. Sasagawa, and T. Tamegai, Zero-energy vortex bound state in the superconducting topological surface state of Fe(Se, Te). Nat. Mater. **18**, 811 (2019).

(6) N. Zaki, G. Gu, A. Tsvelik, C. Wu, and P. D. Johnson, Time-reversal symmetry breaking in the Fe-chalcogenide superconductors, Proc. Nat. Acad. Sci. USA **118**, e200724118 (2021).

(7) N. J. McLaughlin, H. Wang, M. Huang, E. Lee-Wong, L. Hu, H. Lu, G. Q. Yan, G. Gu. C. Wu, Y.-Z. You, and C. R. Du, Strong Correlation Between Superconductivity and Ferromagnetism in an Fe-Chalcogenide Superconductor, Nano Lett. **21**, 7277 (2021).

(8) C. Farhang, N. Zaki, J. Wang, G. Gu, P. D. Johnson, and J. Xia, Revealing the Origin of Time-Reversal Symmetry Breaking in Fe-Chalcogenide Superconductor FeTe$_{1-x}$Se$_x$, Phys. Rev. Lett. **130**, 046702 (2023).

(9) M. Sigrist and K. Ueda, Phenomenological theory of unconventional superconductivity. *Rev. Mod. Phys.* **63**, 239 (1991).
12

**Figures and figure captions**

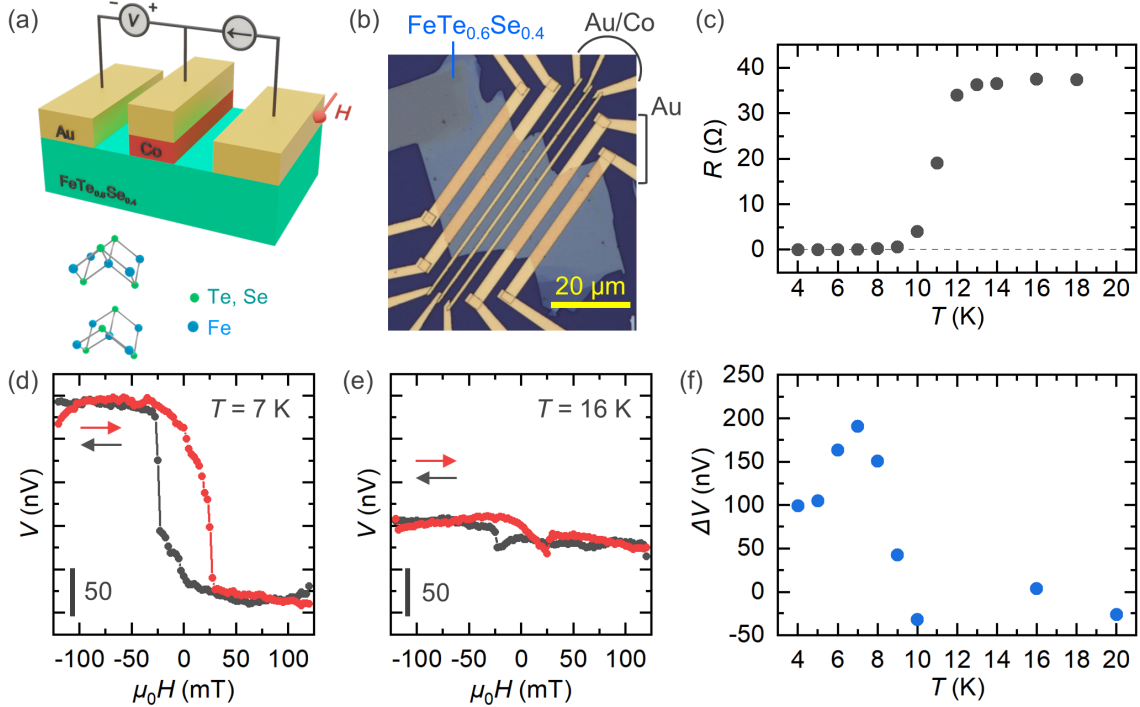

**Figure 1.** (a) Schematic structure of an FeTe$_{0.6}$Se$_{0.4}$ spin device. An electric current is injected from one Au electrode into the Co electrode, exciting spin-polarized Bogoliubov quasiparticles, and spin voltages ascribed to relative spin alignments between the Co and quasiparticles are detected in the voltage circuit. The structure of FeTe$_{0.6}$Se$_{0.4}$ is also shown in the figure. (b) Optical microscopy image of the device. (c) Temperature dependence of the resistance of FeTe$_{0.6}$Se$_{0.4}$. (d) Measured spin voltages in the FeTe$_{0.6}$Se$_{0.4}$ device at 7 K. The amplitude of the spin voltage $\Delta V$ is defined as the difference in the signals between upward and downward sweeping of the external magnetic field. The signals under upward and downward sweeping are shown as solid red and black lines, respectively. (e) Measured spin voltages in the FeTe$_{0.6}$Se$_{0.4}$ device at 16 K, where the anisotropic magnetoresistance due to the Co electrode is dominant. (f) Temperature dependence of $\Delta V$, where the magnitude of $\Delta V$ is the maximum when the superconducting gap opens (8 K, the temperature of zero resistance) and monotonically decreases as the gap expands.

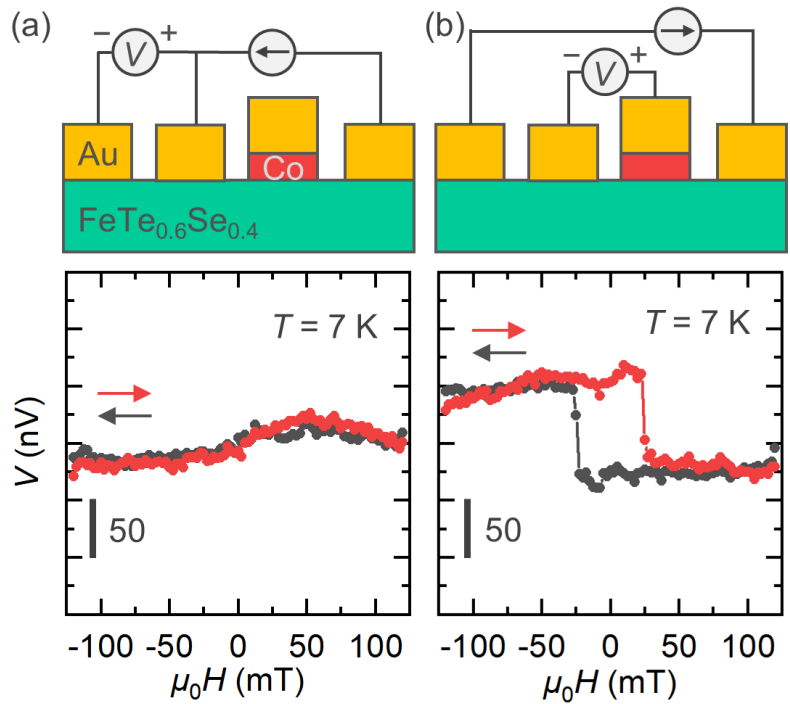

**Figure 2.** (a) Schematic of the three-terminal measurement circuit using a nonmagnetic Au electrode and measured signals at 7 K, below the superconducting transition. (b) Schematic of the four-terminal measurement circuit using the Co electrode. The spin voltage hysteresis is prominent. The scales of the figures showing spin voltage results are set to be the same.

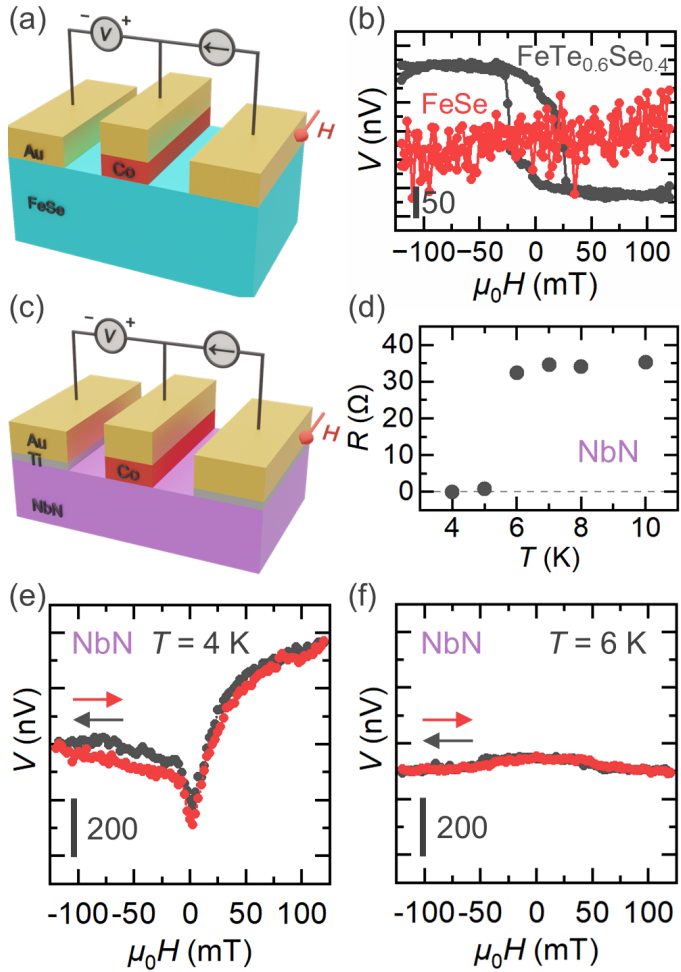

**Figure 3.** (a) Schematic structure of a FeSe spin device. (b) Comparison of spin signals in the FeSe device at 7 K and in the FeTe$_{0.6}$Se$_{0.4}$ devices at 10 K. We note that the signal from the FeTe$_{0.6}$Se$_{0.4}$ device is that shown in Fig. 1(d) and the difference of the measuring temperature is due to the difference in the superconducting transition temperatures. The spin voltage hysteresis appears only in the FeTe$_{0.6}$Se$_{0.4}$ device. (c) a NbN spin device. (d) Temperature dependence of the resistance of NbN. (e) and (f) Measured spin voltages in the NbN device at 4 K, below the superconducting transition temperature of NbN (e), and at 6 K, above the transition temperature (f). No spin voltage hysteresis can be seen in either measurement.

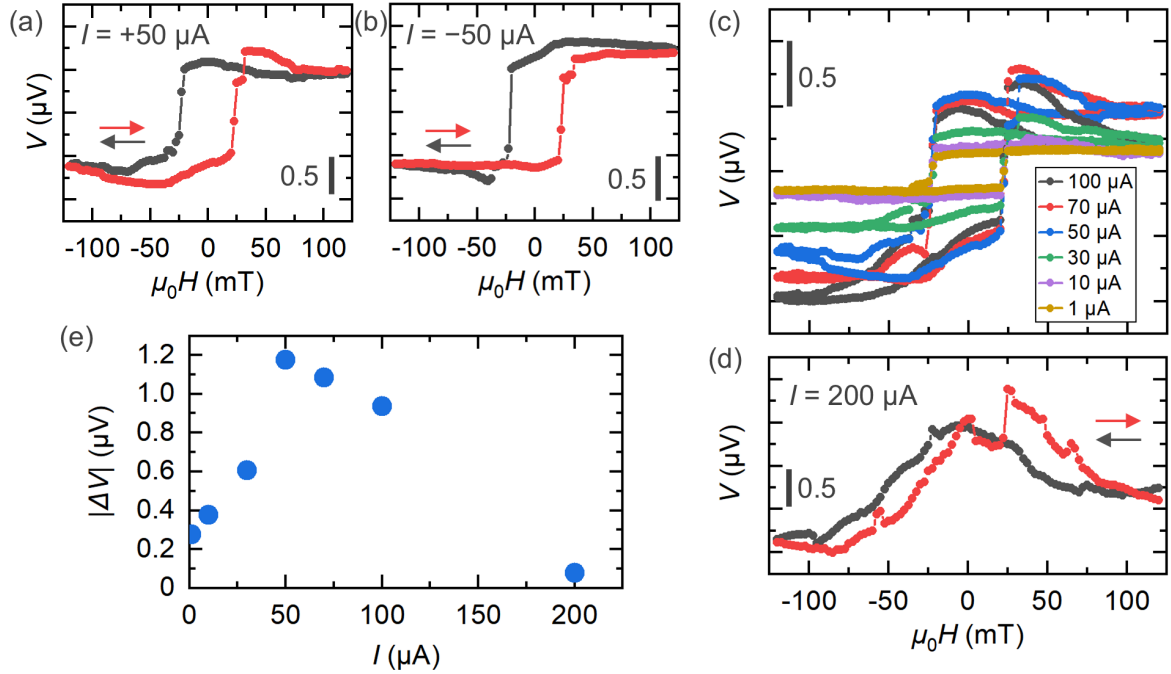

**Figure 4.** (a) and (b) Spin voltages under electric currents $I$ of $+50$ μA (a) and $-50$ μA (b). The polarity of the hysteresis is independent of the current polarity. The AMR signals ascribed to the magnetization reversal of the Co electrode appear at approximately $\pm 25$ mT. (c) Current amplitude dependence of the spin voltage. To remove the offset voltage to allow direct comparison of the amplitudes, half of the difference in the spin voltages under downward and upward sweeping at 0 mT was subtracted. (d) Spin voltage at $I = +200$ μA. The hysteresis disappears because of the breaking of the superconducting state of FeTe$_{0.6}$Se$_{0.4}$. (e) Current amplitude dependence of the spin voltages $\Delta V$. The details of the overall trend are discussed in the main text.